\documentclass{icrc29}
\usepackage{graphicx,amssymb,amsmath,times}
\begin{document}
\title{The trigger system of the Pierre Auger Surface Detector: operation, efficiency and stability}

\author[I.Lhenry-Yvon et al.] { D. Allard, E. Armengaud, I. Allekotte, P. Allison, J. Aublin, M. Ave, P. Bauleo, J. Beatty,
\newauthor 
 T. Beau,  X. Bertou, P. Billoir, C. Bonifazi, A. Chou, J. Chye, S. Dagoret-Campagne, 
\newauthor 
A. Dorofeev, P.L. Ghia, M. G\'omez Berisso, A. Gorgi, J.C. Hamilton, J. Harton,
\newauthor
 R. Knapik, C. Lachaud, I. Lhenry-Yvon, A. Letessier-Selvon, J. Matthews, C. Medina,
\newauthor 
 R. Meyhandan, G. Navarra, D. Nitz, E.Parizot, B. Revenu, Z. Szadkowski, T. Yamamoto  
\newauthor 
for the Pierre Auger Collaboration$^a$ \\
  (a) Pierre Auger Observatory, Av San Mart{\'\i}n Norte 304,(5613) Malarg\"ue, Argentina
}

\presenter{Presenter: I. Lhenry-Yvon (lhenry@ipno.in2p3.fr),  
\ usa-bauleo-PM-abs1-he14-poster}

\maketitle

\begin{abstract}

The trigger system of the Surface Detector (SD) of the Pierre Auger Observatory is described, from the identification of candidate showers ($E>1$~EeV) at the level of a single station, among a huge background (mainly single muons), up to the selection of real events and the rejection of random coincidences at a higher central trigger level (including the reconstruction accuracy). Efficiency of a single station trigger is evaluated using real data , and the  high performance of event selection hierarchy  will be demonstrated. 
\end{abstract}
\vspace{-5mm}
\section{Introduction}
The Pierre Auger Surface Array will consist of 1600 Water Cherenkov detectors sampling ground particles of atmospheric air showers produced by a single energetic particle. The Cherenkov light detected is read out by three large photomultipliers and finally digitized at 40 MHz by Flash Analog Digital Converters (FADC). The detector is extensively described in these proceedings \cite{be05}. The trigger system has been designed to allow the SD of Auger to operate at a wide range of primary energies, for both vertical and very inclined showers with full efficiency for cosmic rays above $10^{19}$~eV.  It should select events of interest and reject background or uninteresting events, while keeping the rate constraints imposed by the communication and data acquisition system. The trigger for the Surface Detector is hierarchical with local triggers at levels 1 and 2 (called T1 and T2), whereas level 3 (T3) is formed at the observatory campus based upon the spatial and temporal correlation of the level 2 triggers.  All data satisfying the T3 trigger are stored.  Additional level of triggers are implemented offline in order to select physical events (T4 physics trigger) and accurate events (T5 quality trigger), with the core inside the array. In section 2 the two levels of the local trigger are described. The efficiency is derived from real data using two different methods. Section 3 is devoted to the selection of physics events (T3 and T4) and the efficiency of the experiment is discussed. In section 4 the quality trigger adopted is presented.
\vspace{-5mm}
\section{Local triggers characteristics}
\begin{figure}[h]
\begin{center}
\includegraphics*[width=0.6\textwidth,angle=0,clip]{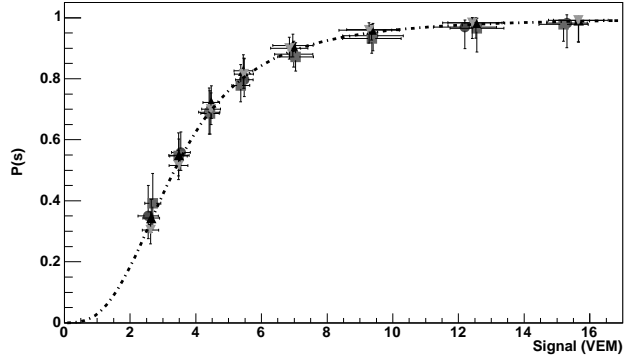}
\caption{\label {fig1} The points represent the two pairs of stations from method one with statistical error bars. For method 1 all showers with $\theta$ less then 60 degrees and S(1000) greater than 2 VEM have been used.  The dotted line is the fit from method 2 using the same large theta and S(1000) bins.
The functional form of the fit is $P(S) = s^{N}/(s^{N}+ s_{50\%}^{N}).$}
\end{center}
\end{figure}

Two different trigger modes are currently implemented at the T1 level. The first uses a Time over Threshold (ToT) trigger, requiring that 13 bins in a 120 bins window are above a threshold of 0.2 $I_{VEM}^{est}$ in coincidence on 2 PMTs \cite{ni01}. The estimated current for a Vertical Equivalent Muon ($I_{VEM}^{est}$) is the reference unit for the calibration of FADC traces signals \cite{al05}. This  trigger has a relatively low rate of about 1.6~Hz, which is the expected rate for double muons for an Auger tank. It is extremely efficient for selecting small  but spread signals, typical for high energy distant EAS or for low energy showers, while ignoring single muons background. The second trigger is a 3-fold coincidence of a simple 1.75 $I_{VEM}^{est}$ threshold. This trigger is more noisy, with  a rate of about 100 Hz, but it is needed to detect fast signals ($<$~200 ns) corresponding to the muonic component generated by horizontal showers. \\
The T2 trigger is applied in the station controller to select from the T1 signals those  likely to have come from EAS and to reduce to about 20 Hz the rate of events to be sent to the central station. 
All ToT triggers are directly promoted T2 whereas  T1 threshold triggers are requested to pass a higher threshold of 3.2 $I_{VEM}^{est}$ in coincidence for 3 PMTs. Only T2 triggers are used for the definition of a T3.\\
\hspace*{5mm}
The probability for  a station to pass the trigger requirements strongly depends on the integrated signal.  We define this probability, P(S), as the ratio of stations that trigger divided by the number of stations for a given integrated signal. We have measured this probability directly from the data by two different methods.
The first method is based on the existence of two pairs of detectors
separated by 11 m from each other.  The double sampling of signals in
near locations provides a way to estimate the number of signals that
did not cause a trigger.  This method allows direct comparison of P(s)
for individual stations. In figure \ref{fig1} we show that the two
pairs of  stations used in this study have the same P(s) within
uncertainties. The second method uses this result and assumes that a
single P(s) can describe the behavior of all the stations in the
surface array.  For each event a Lateral Distribution Function (LDF)
is fitted to all the stations that have signal. Here the LDF used  is a
parabola on a log-log scale. The systematic due to the use of a LDF 
form different from the one described in this conference is of a few percent, within the statistical uncertainties 
of the method \cite{ba05}.  The LDF is assumed to be cylindrically symmetric so for each event there are regions of constant signal. In each event the stations that did and did not trigger in a given constant signal region can be identified and P(s) computed.
The agreement between both methods is shown in figure 1.  Method one has low statistics so any further dependencies  on the trigger probability can not be identified.  Method two has high statistics and can further parameterize P(s) into zenith angle ($\theta$) and shower size parameter S(1000) bins.
Correct knowledge of the trigger probability is needed for the
acceptance estimation \cite{pa05}. \\
\hspace*{5mm} 
The stability of the trigger rates is of great importance for a good estimation of the acceptance of the array. 
The threshold T2 rates are uniform over the present array within a few
percent.  The ToT T2rates are more spread, since they are sensitive to
the charge of the signal that depends on the characteristics of the
water in the tank. The decay time of the pulses is a good estimator of
the water quality in the tank. It has been shown \cite{he05} that each tank needs a few months after installation where this decay time slowly decreases then stabilizes to an average value around 65 ns. Once the tanks are stable, the average ToT rate over the array is 1.6 $\pm1$~Hz.  The ToT rate is also dependent on temperature, its variation is carefully studied \cite{be05} and the influence on the higher level triggers is shown in the next section.
\vspace{-5mm}
\section{Event selection}
The main Auger T3 trigger requires  the coincidence of 3 tanks which have passed the ToT conditions and meeting the requirement of a minimum of compactness (one of the tanks must have one of its closest neighbors and one of its second closest neighbors triggered). Since the ToT as a local trigger has already very low background (mainly double muons), this so-called 3ToT trigger selects mostly physical events. The rate of this T3  with the present  number of working detectors in the array is around 600 events per day, or 1.3 events per triangle of 3 neighboring working stations . This trigger is extremely relevant since 90\% of the selected events are showers and is mostly efficient for vertical showers. The other implemented trigger is more permissive.  It requires a four-fold coincidence  of any T2 with a moderate compactness requirement (among the 4 fired tanks, one can be as far as 6 km away from others within appropriate time window). Such a trigger is absolutely needed  to allow for the detection of horizontal showers that generate fast signals and have wide-spread topological patterns. This trigger selects about 400 events per day, but only 2\% are real showers. \\
\hspace*{5mm}
 A physical trigger (T4) is needed to select only showers from the set of stored T3 data. An official physical trigger is applied offline to select events for zenith angles below 60 degrees. The chosen criteria use two main characteristics of vertical showers. The first one is the compactness of the triggered tanks, the second one is the fact that most  FADC traces  are spread enough in time to satisfy a ToT condition. It was shown that requiring a 3 ToT compact configuration in an event ensures that more than 99\% are showers. The present physics trigger is dual and requires either a compact 3 ToT or a compact configuration of any local trigger called 4C1 (at least one fired station has 3 triggered tanks out of its 6 first neighbours). The tanks satisfying the 3 ToT or 4C1 condition must have their trigger time compatible with the speed of light (with a tolerance of 200~ns to keep very horizontal rare events).  With the 3ToT T4 trigger, less than 5\% of showers below 60 degrees are lost. The 4C1 trigger , whose event rate is about 2\% of the previous one , ensures to keep the 5\% of the showers below 60 degrees lost by the other T4 and also selects low energy events above 60 degrees. This can be seen in figure 2 where  is shown on the left  the distribution in angle of events selected by the different T4s and on the right the energy distribution.
\begin{figure}[h]
\begin{center}
\includegraphics*[width=.95\textwidth,angle=0,clip]{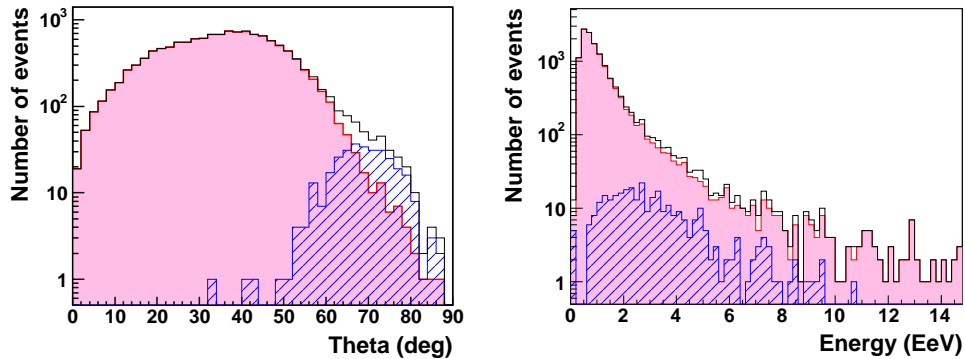}
\caption{\label {fig2} Zenith angle (left plot) and energy (right plot) distribution for events satisfying the adopted T4. The red shaded area correspond to events passing 3ToT and the blue dashed area to events passing the 4C1. Those that are both 3ToT and 4C1 are counted in the 3ToT distribution}
\end{center}
\end{figure}
In most selected events, a number of accidental tanks need to be removed. A complete procedure has been defined to fulfill this task. It is based on the definition of a seed that is an elementary triangle (one station with 2 neighbors in a non-aligned configuration).  The seed with the highest total signal is used to define a plane front of the shower. Then all stations are examined, and are defined as accidental if their time delay to the front plane propagation is outside a defined time window. Isolated tanks are also removed. For the period of March 2005 for example, the average number of discarded tanks per event is 4 which is compatible with the expected number of accidental tanks for the corresponding number of working tanks, $N_{\mathrm{total}}$, where \\ $N_{\mathrm{accidental}} =N_{\mathrm{total}} *\mathrm{ T1\_rate}* \mathrm{T3\_timewindow} = 667 * 100\mathrm{Hz}* 60~\mu \mathrm{s} = 3.96$ per event.\\
99\% of the obtained events are reconstructed. The definition of a T4 for horizontal showers is more challenging and is still under study \cite{ne05}. The effect of the temperature on the T4 rate was studied. The number of T4 events per day divided by the number of active triangles of stations has a dependence on temperature of about 1\% per degree. This  has to be taken into account for the estimation of the acceptance of the array when it is not saturated .
\vspace{-5mm}
 \section{T5 quality Trigger}
One further step is needed to compute the acceptance of the detector and build the spectrum. Among the events having passed the T4 trigger, only those that can be reconstructed with a known energy and angular accuracy will be used. This is the task of the T5 quality trigger. Various studies have been performed to identify under which conditions events could satisfy this requirement. The main problem to solve is that if an event is close to the border of the array, a part of the shower is probably missing, and the real core could be outside of the existing array, whereas the reconstructed core will be by construction inside the array, in particular for low multiplicity events. Such events will have wrong core positions, so wrong energies and typically should not pass the quality trigger. \\
\hspace*{5mm}
 The adopted T5 requires that the tank with highest signal must have at least 5 working tanks among its 6 closest neighbors at the time of the event and more over, the reconstructed core must be inside an equilateral triangle of working stations. This  represents an efficient quality cut by guaranteeing that no crucial  information is missed for the shower reconstruction. This study  described in \cite{gh05} evaluates the effect of T5 on the accuracy of the reconstruction and in particular on the signal at 1000 m from shower axis S(1000). The maximum systematic uncertainty  in the reconstructed S(1000) due to event sampling into the array or to the effect of a missing internal tank is around 8\%.
\vspace{-5mm}
\section{Conclusion}
The hierarchical trigger of the Pierre Auger Observatory has been fully described. This trigger chain allows to decrease the event rate in a single tank from 3 kHz due to mainly background muons up to 3 per day, due to real showers, corresponding to a rejection factor of $10^{8}$.


\vspace{-5mm}
\begin{thebibliography}{99}
\bibitem {be05} Pierre Auger Collaboration, these proceedings, arg-bertou-X-abs1-he14-oral. 
\bibitem {ni01} D. Nitz for the Pierre Auger Collaboration, \ ICRC 2001. 
\bibitem {al05} M. Aglietta {\it et al}, these proceedings, usa-allison-PS-abs1-he14-poster. 
\bibitem {ba05} D. Barnhill {\it et al}, these proceedings usa-bauleo-PM-abs2-he14-poster. 
\bibitem {pa05} D. Allard {\it et al}, these proceedings fra-parizot-E-abs1-he14-poster. 
\bibitem {he05} I. Allekotte {\it et al}, these proceedings, usa-arisaka-K-abs1-he15-poster. 
\bibitem {ne05} Pierre Auger Collaboration, these proceedings,mex-nellen-L-abs1-he14-oral. 
\bibitem {gh05} Pierre Auger Collaboration, these proceedings, ita-ghia-P-abs1-he14-oral. 
\bibitem {so05} Pierre Auger Collaboration,these proceedings,usa-sommers-P-he14-oral. 
\end{thebibliography}
\end{document}